\documentclass[12pt]{article}
\usepackage{graphicx}
\usepackage{amssymb}
\usepackage{amsmath}
\usepackage{array}
\usepackage{enumerate}
\usepackage{psfrag}

\advance \topmargin by -\headheight
\advance \topmargin by -\headsep
\evensidemargin \oddsidemargin
\def\mmu{$\mu$} 
 
\def\ai{$a_{i}$} 
\def\xsubibar{$\bar{x}_{i}$} 
\def\xsubi{$x_{i}$} 
\def\xfour{$x_{4}$} 
\def\xmu{$x_{\mu}$} 
\def\pmu{$p_{\mu}$} 
\def\gammamu{$\gamma^{\mu}$} 
\def\psubi{$p_i$} 
\def\alphai{$\alpha_{i}$} 
\def\betai{$\beta$} 
\def\etai{$\eta_{i}$} 
\def\etafour{$\eta_{4}$} 
\def\etamu{$\eta_{\mu}$} 
\def\hone{$H_{1}$} 
\def\htwo{$H_{2}$} 
\def\honetwo{$H_{12}$} 
\def\zone{$z_{1}$} 
\def\ztwo{$z_{2}$} 
\def\xiit{$\xi_i(t)$} 
\def\xii{$\xi_i$} 
\def\ui{$u_{i}$} 
\def\uibar{$\bar{u}_{i}$} 
\def\umu{$u_{\mu}$} 
\def\umubar{$\bar{u}_{\mu}$} 
\def\ufour{$u_{4}$} 
\def\ufourbar{$\bar{u}_{4}$} 
\def\xbar{$\bar{x}$} 
\def\tbar{$\bar{t}$} 
\def\duidt{$\frac{du_i}{dt}$} 
\def\duidtbar{$\frac{du_i}{d\bar{t}}$} 
\def\duibardt{$\frac{d\bar{u}_i}{dt}$} 
\def\duibardtbar{$\frac{d\bar{u}_i}{d\bar{t}}$} 
\def\pone{$p_{1}$} 
\def\lone{$L_{1}$} 
\def\ltwo{$L_{2}$} 
\def\lthree{$L_{3}$} 
\def\lfour{$L_{4}$} 
\def\psione{$\psi_1$}
\def\psitwo{$\psi_2$}
\def\psithreefour{$\psi_{34}$}

\begin{document} 
\thispagestyle{empty}

{\bf Extension of Dirac theory and the classification of elementary
particles}

Janet Pan$^1$ and Lu Lin$^2$

1. Yale University, P.O. Box 208284, New Haven, CT 06520-8284.

2. Wufeng Institute of Technology, Chia-yi, Taiwan, ROC.
E-mail : lulin@mail.wfc.edu.tw

{\bf Abstract}

The Dirac theory implies the existence of an internal vector space, in
addition to spin space.  Using Dirac's coupling of variables in internal
space to those in physical space, we construct a new configuration
structure for particles in the combined physical plus internal spaces.  The
importance of this is that the internal degrees of freedom implicit in
Dirac's theory allow a new classification of elementary particles.  An
important consequence is the prediction of a new type of quark.  As
expected, our theory groups fermions into doublets, which are then divided
into color singlets (leptons) and color triplets (quarks), and which are
then further divided into generation singlets and generation triplets.  If
the Pauli exclusion principle for fermions is also valid within a
particle's internal space, then our theory makes two important
predictions.  First, we can explain why the widely studied quarks (up,
down, charm, strange, top, bottom) cannot be observed as free states in
nature.  Second, we predict the existence of a new quark which can indeed
be observed as a free state in nature, and whose wave function is
antisymmetric in internal space.  W.~M. Fairbank, a Guggenheim Fellow, has
published experimental data which supports our second prediction.

\newpage

Consider the motion of a free electron (or, in general, a
fermion) described by a space-time 4-vector \xmu, its momentum \pmu,
and Dirac matrices \alphai, \betai, (i=1 to 3, and \mmu=1 to 4). The
Dirac Hamiltonian \cite{bethe1,dirac1} of this system is 
\begin{equation}
H=\sum\alpha_i\cdot p_i + \beta m, \quad\text{where }
\hbar = c =1.
\label{eq:DiracH}
\end{equation}
The i-th component of the velocity and the i-th component of the
coordinate are given as (see Chapter XI of ref. \cite{dirac1}) 
\begin{equation}
\dot{x}_i = \alpha_i = \dot{\xi}_i +\dot{\eta}_i
\label{eq:xidot}
\end{equation}
\begin{equation}
x_i = \xi_i +\eta_i
\label{eq:xi}
\end{equation}
\begin{equation}
\xi_i = a_i + p_i H^{-1} t
\label{eq:xii}
\end{equation}
\begin{equation}
\eta_i  = \frac{i}{2} \left(\alpha_i H^{-1} - p_i H^{-2} \right) 
\label{eq:etai}
\end{equation}
where \ai\ is the \xiit\ at t=0. Since \alphai\ does not commute with the
Hamiltonian, each component of the velocity of the electron is not a
constant of the motion.  Also, since
\alphai\ has eigenvalues $\pm$1, then a measurement of any
component of the velocity of the electron will give the speed of light,
even when \psubi\ is zero. On the other hand, since the square of the
velocity operator is \mbox{$v^2=\sum\alpha_i^2=3c^2$}, which is a constant,
independent of all \psubi, then \mbox{$v^2$} commutes with
the Hamiltonian and can be simultaneously measured with the Hamiltonian.
However, the result, \mbox{$v^2=3c^2$}, violates relativity, and forces the
electron mass \mbox{$m(v)=m\left(1-\frac{v^2}{c^2}\right)^{-\frac{1}{2}}$}
to have an imaginary value, even when the momentum \psubi\ is zero. This
does not make sense. This trouble can be understood and reinterpreted as the
following.  According to Dirac \cite{dirac1}, since the \alphai\ and the
\betai\ are independent of, and commute with \xmu\ and \pmu, then they must
describe some new degree of freedom, belonging to some internal motion of
the electron.  This implies the existence of an internal space,
in which the electron moves, and which is independent of physical space
(PS).  The Hamiltonian H in Equation.~(\ref{eq:DiracH}) requires that the
motion of the electron in physical space is coupled to its motion in 
internal space.  We can assert that the same electron is moving in an
enlarged space, which is comprised of physical space plus an internal
space. Another way to look at the problem is that the particle energy has
two contributions, and these two contributions are coupled by the Dirac
Hamiltonian. This is analogous to the case of two classical oscillators,
with coordinates \zone\ and \ztwo, and Hamiltonians \hone\ and \htwo. If the
oscillators are coupled, then the total Hamiltonian will be
H=\hone+\htwo+\honetwo, and the normal coordinates will be functions of
\zone\ and \ztwo.

Consider an operator \mbox{$c^2p_iH^{-1}$}, which commutes with the
Hamiltonian H and has an eigenvalue \mbox{$\frac{p_i}{M}$}, where M the
total mass of the electron. This value is just the value of
\mbox{$v_i=\dot{\xi}_i$} in Equations (\ref{eq:xidot}) and (\ref{eq:xii}),
and is the usual velocity of the electron. Thus, we can interpret 
\mbox{$x_i=\xi_i+\eta_i$} as a generalized coordinate vector, with
\xii\ in  physical space and \etai\ (which is, in general, complex) in
internal space.  Then, \mbox{$\dot{x}_i=c\alpha_i$} is the velocity of the
generalized coordinates, while \mbox{$v_i=\dot{\xi}_i$} is the physical
velocity of the electron.  (The motion of \etai\ is known as Zitterbewegung
\cite{bethe1} and is unphysical.)

Since \mbox{\gammamu\pmu=m} is a scalar operator obtained from
two 4-vectors, where \gammamu\ is another representation of \alphai, \betai,
then this implies that the internal space is 4-dimensional, and a vector in
this space has 4 independent components.  Thus, \etai\ should have a 4th
component, \etafour, to make a 4-vector \etamu.  Now we can define our total
space as the greater space (GS), which is comprised of a 4-dimesional
physical space (PS) and a 4-dimensional internal space (IS). A vector \xmu\
in PS has real \xsubi\ and imaginary \xfour.  However, we do not have any
direct information about the nature of a vector in IS.  We now proceed to
study the nature of a vector in IS. 

Denote the vectors in PS, and in IS by \umu\ and \umubar\ respectively,
and the time variables corresponding to \ufour\ and \ufourbar\ by t and
\tbar. The Lagrangian of our system can be generalized to contain variables
of both PS and IS. For a free particle in classical mechanics, the
Lagrangian is a scalar, and is a function of the square of the velocity of
the particle. Velocity terms like \duibardt\ appear in Equation
(\ref{eq:xidot}), which implies that, by symmetry, terms like \duidtbar\
should be included in the generalized Lagrangian.  This generalized
Lagrangian should contain all the velocity terms,
\duidt, \duidtbar, \duibardt, and \duibardtbar, plus small coupling terms.
Since we are interested only in the main qualitative features of the
Lagrangian, in order to elucidate the nature of the 4-vector in IS, then we
may neglect the small coupling terms and express the Lagrangian as
\begin{eqnarray}
L & = 
L_1\left(\frac{du_i}{dt}\right)+
L_2\left(\frac{du_i}{d\bar{t}}\right)+
L_3\left(\frac{d\bar{u}_i}{dt}\right)+
L_4\left(\frac{d\bar{u}_i}{d\bar{t}}\right)
\\ 
& = 
L_1\left(u_i,t\right)+
L_2\left(u_i,\bar{t}\right)+
L_3\left(\bar{u}_i,t\right)+
L_4\left(\bar{u}_i,\bar{t}\right)
\label{eq:sumL1toL4}
\end{eqnarray}
This Lagrangian now allows the system to be treated like a problem with
four independent particles. \lone\ describes a physical particle in PS.
Each of the other three parts of the Lagrangian has either an unphysical
particle or an unphysical space. \ltwo\ describes a particle in PS, but
which is varying with respect to its internal time.  The particle described
by \ltwo\ has coordinate variables which should be considered to be
independent of the \ui\ in \lone.  Since the motion described by \ltwo\ is
unphysical, then the  location described by \ltwo\ should be imaginary.
Furthermore, the internal time carried by an unphysical real particle
should also be considered as imaginary. For \lthree\ and \lfour, we can
assume that, when the universe was created, the four vectors were created
together by taking all possible real and imaginary combinations of the
(\ui,t)  with equal probability.   As a result, we have the greater space 
\begin{eqnarray}
GS & = 
P_1\left(u_i,t\right)+
P_2\left(iu_i,it\right)+
P_3\left(\bar{u}_i,i\bar{t}\right)+
P_4\left(i\bar{u}_i,\bar{t}\right)
\label{eq:GSP1toP4}
\end{eqnarray}
where \ui, \uibar, t, \tbar\ are real numbers. In
1998, one of the authors \cite{lin1} suggested this simple idea of a complex
space-time, and now we develop it further. 

In physical space, if we can describe the
system of \mbox{\lone+\ltwo} of Equation~(\ref{eq:sumL1toL4}), then
a well-behaved wave function  \mbox{$\Phi(x_i,i\bar{x}_i;t,i\bar{t})$} must
exist. For a stationary state, the spatial distribution must be independent
of time, so $\Phi$ can be factored,
\begin{equation}
\Phi=\Phi(X)\Phi(T),
\quad\text{where }X=x+i\bar{x}\text{ and }
T=t+i\bar{t}
\end{equation}
We can reasonably assume the simple conditions that
$\Phi$ is a harmonic function of \mbox{$(x,i\bar{x})$} and
\mbox{$(t,i\bar{t})$} respectively, and that $\Phi$  has continuous partial
derivatives of second order which satisfy the Laplace equations 
\begin{eqnarray}
\frac{\partial^2\Phi(X)}{\partial x^2} +
\frac{\partial^2\Phi(X)}{\partial\bar{x}^2}& =0
\label{eq:LaplaceX}
\\
\frac{\partial^2\Phi(T)}{\partial t^2} +
\frac{\partial^2\Phi(T)}{\partial\bar{t}^2}& =0
\label{eq:LaplaceT}
\end{eqnarray}
Since $x$ and $i\bar{x}$ are independent, we can justifiably write the
factorization \mbox{$\Phi(X)=\Phi(x)\Phi(i\bar{x})$} , and similarly
for \mbox{$\Phi(T)$}. The solutions can be written as
\begin{eqnarray}
\Phi(X) &= A\exp(\pm ikx)\exp(\pm ik(i\bar{x}))
\label{eq:PhiX}
\\
\Phi(T) &= B\exp(-i\omega t)\exp(-i\omega(i\bar{t}))
\label{eq:PhiT}
\end{eqnarray}
In Equation~(\ref{eq:PhiX}) we have chosen the direction of $k$ along the
$x$ axes, and the $i\bar{x}$ along the x-axes but independent of $x$. In 
Equation~(\ref{eq:PhiT}), note that we are concerned with a
free electron, which is not an anti-particle, and therefore we must take the
positive frequency solution for the wave function. This is also true for
wave functions in \ltwo, \lthree, \lfour, because they describe the motion
of electrons in different regions of greater space. If we had
taken both the positive and the negative frequencies for $\omega$, then the
final total wave function would have the form of \mbox{$a\exp(-i\omega
t)+b\exp(i\omega t)$}, which is a mixture of a particle and an
anti-particle. The latter should be clear because negative frequency
solutions correspond to anti-particles.  This is easily seen by
transforming the Dirac equation in a electromagnetic field by a
time-reversal plus a charge conjugation. However, since the charge operator
commutes with the Hamiltonian, then the charge eigenvalue can be
simultaneously measured with the energy eigenvalue. Thus, the energy
eigenfunction must also be a charge eigenfunction.  Hence, energy
eigenfunctions are not comprised of mixed charge states.  That is to say,
energy eigenfunctions are not comprised of a mixture of particle and
anti-particle states.  By setting \mbox{\xbar=\tbar=0}, we get a projected
free wave solution in physical space with only real space-time variables,
corresponding to a particle with energy and momentum as
\mbox{$E=\hbar\omega$} and \mbox{$p=\hbar k$}. 

Consider the four independent particles specified by the four 4-vectors of
equation~(\ref{eq:GSP1toP4}). The single particle states are generated by
solving, for each particle, a Dirac equation in the corresponding region.
In the physical \xsubi-space, there are two single particles, a physical
particle and an unphysical particle, while there are two identical
particles, both unphysical, in the internal \xsubibar-space.  For a
stationary state, a total state function is the product
\mbox{\psione\psitwo\psithreefour}, where \psione\ and \psitwo\ are single
particle states for particles 1 and 2, and \psithreefour\ is a two-particle
state for particles 3 and 4. Each
\mbox{$\psi$} has a spatial
\xsubi-part and a spin part. The \xsubi-part of each \mbox{$\psi$} must satisfy
equation~(\ref{eq:LaplaceX}). The solution in equation~(\ref{eq:PhiX}) is
the \xsubi-parts of \psione\psitwo. Since \pone\ corresponds to physical
space, \psione\ must have a good momentum, so there is only one momentum
state in \psione.  \psitwo\ can take two momentum states, those in
equation~(\ref{eq:LaplaceX}). Similarly, for the unphysical regions of
\mbox{\lthree+\lfour}, we have \mbox{$2\bigotimes 2$} momentum states. For
the spin part of the wave function, since the spin is diagonalized in the
rest frame of \lone, the particle can only take one spin state in \lone.
This state cannot be mixed with any unphysical spin state. In the other
three parts,  \ltwo, \lthree, \lfour\ of unphysical space, each part has
two independent spin states. Therefore, for a fermion, with spin states and
momentum states taken together, the total number of independent internal
states generated from
\mbox{\ltwo+\lthree+\lfour} is 
\begin{eqnarray}
\text{Number of internal states} 
& = & 2\bigotimes 2\bigotimes 2\bigotimes2\bigotimes 2\bigotimes 2\\
& = & 2\bigotimes\left[1\bigoplus3\right]\bigotimes\left[1\bigoplus 3\right]
\bigotimes 2
\label{eq:NB}
\end{eqnarray}
The first Ô2Õ and the last Ô2Õ in Equation~(\ref{eq:NB}) come from
the spin states and the momentum states of \ltwo\ while the two brackets
come from those states of \lthree\ and \lfour. The first Ô2Õ in
Equation~(\ref{eq:NB}) means that, all fermions are grouped in doublets,
such as \mbox{$(\nu e),(\nu_{\mu}\mu),(\nu_t t)$} and
\mbox{$(ud),(cs),(tb)$}. The first bracket in Equation~(\ref{eq:NB})
signifies that, all the above doublets are divided into color singlets
with zero color plus color triplets (each quark has three color states). The
second bracket signifies that, the above entities are further divided into a
generation singlet which we shall explain later, plus a generation triplet
which contain three lepton doublets \mbox{$(\nu e),(\nu_{\mu}\mu),(\nu_t
t)$} and three quark doublets \mbox{$(ud),(cs),(tb)$}. As far as
the last Ô2Õ is concerned, we postulate that when our universe was
created, the creator might have designed a duality for particles to have a
symmetry of electricity and magnetism, such as an electron and a magnetron,
a proton and a magnetic proton, a quark and a magnetic quark. However,
no experimental evidence exists for magnetic charges,
or monopoles.  Thus far, experimental evidence is consistent only with
asymmetric Maxwell equations: that we have
\mbox{$\nabla\cdot E =4\pi\rho$}, but \mbox{$\nabla\cdot B=0$}. Could
the magnetic matter exist elsewhere in the universe?  If it could, then
this would vindicate the brilliant theoretical argument of Dirac
\cite{dirac2}: that the mere existence of one magnetic monopole in the
universe would offer an explanation of the discrete nature of electric
charge, and thus solve one of the most profound mysteries of the physical
world. 

For a one particle state, define the color-spin and the generation-spin as
\mbox{$(s_c,m_c)=(\frac{1}{2},\pm\frac{1}{2})$} and
\mbox{$(s_g,m_g)=(\frac{1}{2},\pm\frac{1}{2})$}, respectively. Then the two
particle color-spin wave functions \mbox{$\Phi_c(SM)$} can be written as a
singlet state \mbox{$\Phi_c(0,0)$}, and triplet states
\mbox{$\Phi_c(1,1)$}, \mbox{$\Phi_c(1,0)$}, \mbox{$\Phi_c(1,-1)$}.
The singlet is anti-symmetric:
\begin{eqnarray}
\Phi_c(0,0) &= 
\frac{1}{\sqrt{2}}
\left(\phi_1^{(+)}\phi_2^{(-)}-\phi_1^{(-)}\phi_2^{(+)}
\right)
\end{eqnarray}
The triplet is symmetric:
\begin{eqnarray}
\Phi_c(1,0) & = 
\frac{1}{\sqrt{2}}
\left(\phi_1^{(+)}\phi_2^{(-)}+\phi_1^{(-)}\phi_2^{(+)}
\right)
\\
\Phi_c(1,1) & =
\phi_1^{(+)}\phi_2^{(+)}
\\
\Phi_c(1,-1) & =
\phi_1^{(-)}\phi_2^{(-)}
\end{eqnarray}
The two particle generation-spin wave functions \mbox{$\Phi_g(S'M')$} can be
written in a similar way.

In the total wave function \psione\psitwo\psithreefour\ of a fermion,
\psione\ and \psitwo\ are single particle wave functions, and \psithreefour\
is a two identical particle wave function in internal space, as was
discussed above. We only need to consider
\mbox{\psithreefour=$\Phi_c(SM)\Phi_g(S'M')$}. We assert now that the
exclusion principle is also valid in internal space, that the total
wave function of a system of identical fermions must be anti-symmetric. All
leptons have singlet color wave functions \mbox{$\Phi_c(0,0)$} with zero
color, which are anti-symmetric. The three generations of leptons,
\mbox{$(\nu e),(\nu_{\mu}\mu),(\nu_t t)$}, have symmetric triplet generation
wave functions,\mbox{$\Phi_g(1,M)$}, \mbox{$M=1,0,-1$}, so their total wave
function \mbox{$\Phi_c(00)\Phi_g(1M)$} are anti-symmetric.
Therefore, the above three doublets of leptons do not violate the exclusion
principle, and are free states in nature. There are also  singlet
generation leptons \mbox{$\Phi_g(0,0)$}, for which the total wave functions 
\mbox{$\Phi_c(00)\Phi_g(00)$} are symmetric. These wave functions violate
the exclusion principle, and cannot be found as free states. In this
work, we shall not discuss the possibility of having compounds states for
this kind of particle. 

All quarks have triplet color wave functions,
\mbox{$\Phi_c(1,M)$}, \mbox{$M=1,0,-1$}, which are symmetric. The three
generations of quarks, \mbox{$(ud),(cs),(tb)$}, have symmetric generation
wave functions, \mbox{$\Phi_g(1,M)$}, so the total wave functions,
\mbox{$\Phi_c(1M)\Phi_g(1M)$} are symmetric. These wave functions also
violate the exclusion principle, and cannot be found as free states. They
can only exist in a compound system. 

There is one last case, where a quark takes triplet color state
\mbox{$\Phi_c(1,M)$} and a singlet generation state \mbox{$\Phi_g(00)$}.
This wave function is anti-symmetric, which does not violate the
exclusion principle, and thus this quark can exist as a free state in
nature. In 1981, W. M. Fairbank, a Guggenheim Fellow, and his colleagues
\cite{fairbank1, fairbank2, fairbank3} reported measurements which showed
unambiguously the existence of fractional charges of
\mbox{$\pm\frac{1}{3}$}. If Fairbank's measurements are correct, then his
quarks could be identified as the quarks which we described as
\mbox{$\Phi_c(1,M)\Phi_g(00)$}. However, other groups need to confirm
Fairbank's rsults, and further work is certainly needed.

\end{document}